\documentclass[final,3p,times, onecolumn]{elsarticle}

\usepackage{latexsym}
\usepackage{bm}

\usepackage{amsmath}
\usepackage{graphicx}   
\usepackage{enumerate}   

\usepackage [latin1]{inputenc}

\usepackage{amssymb}

\begin{document}

\begin{frontmatter}



\title{The congruence of spacelike curves of tachyons with respective energy-momentum tensor of perfect fluid type}

\author{Wytler Cordeiro dos Santos}
\ead{wytler.cordeiro@unb.br}
\address{Universidade de
Bras\'\i lia, CEP 70910-900, DF, Brasil}

\begin{abstract}
The Special Relativity allows the possibility of a class of particles, known tachyons, that have spacelike 4-velocities, i.e., which move with velocity greater than speed of light in vacuum. In this existence frame, the  tachyons  have energy and momentum and they must contribute to the gravitational field through by the means of  the energy-momentum tensor. The superluminal perfect fluid tensor is obtained assuming the framework Lagrangian formalism and spacelike 4-velocities flowlines of tachyons with energy density.  The perfect fluid of tachyons gives rise  a positive  energy density and a form of negative pressure, which according to various cosmological theoretical results, it is associated with dark energy. 
\end{abstract}

\begin{keyword}
  Tachyons, perfect fluid, negative pressure, dark energy  



\end{keyword}

\end{frontmatter}



\section{Introduction}

In the context of Cosmology, the General Relativity states that the matter of the universe composes a  fluid type, named  cosmological fluid, and a cosmological constant $\Lambda$, that  adjust to the cosmological field equations of Friedmann-Lema\^{i}tre. 
Differently of the nature of the normal matter that  yields gravitational attraction, the cosmological constant $\Lambda$ can give rise to a repulsive gravitational force that explains the expansion of the universe discovered by Hubble in 1929  \cite{Cheng}. Meantime by 1998 experimental observations  of distant type Ia superaovae, arguing strongly that the universe has undergone through a phase of accelerated expansion \cite{Riess}. Today it is acceptable that a strange fluid-like entity with a negative pressure often referred to as the dark energy drives the accelerating expansion of the universe \cite{Cheng, Masiero}.

In order to explain the cause of accelerating expansion of the universe, some authors have proposed a hypothetical form of dark energy, a scalar field announced with the name quintessence. Unlike the cosmological constant which, by definition is a constant, the quintessence is dynamic, i.e., it changes over time \cite{Carrol_1,Ratra_1,Caldwell}.
Parallel to the quintessence hypothesis, the hypothesis of the rolling tachyon condensate, in a class of string theories, can be highlighted. Sen have shown that the decay of D-branes produces a pressure-less gas with finite energy density that resembles classical dust \cite{Sami, Sen_1, Sen_2}. Attempts have been made to construct viable cosmological model using rolling tachyon field as a suitable candidate for inflaton or dark energy \cite{Sami,Sadeghi,DeBenedictis,Albergh}.

Some research authors have shown that the Einstein field equations have solutions for a hypothetical tachyon dust or tachyon fluid \cite{Foster, Schwartz_1,Singh, Srivastava}. For this article, it is proposed to write a review on the topic of a tachyon perfect fluid, which in principle can be obtained from a Lagrangian of matter. The energy-momentum tensor of a perfect fluid is calculated by the variation of the action of matter making it come up like this the metric energy-momentum tensor (or Hilbert energy-momentum tensor). So, according to the General Realtivity, the Einstein's field equations states that the curvature of spacetime is related to the distribution of matter represented by the energy-momentum tensor through the equation,
\begin{equation}
\label{Einstein Equation}
 R_{\mu\nu} - \frac{1}{2}Rg_{\mu\nu} + \Lambda g_{\mu\nu} = 8\pi G T_{\mu\nu}.
\end{equation}
$\Lambda$ is cosmological constant, $R_{\mu\nu}$ the contract curvature tensor (Ricci tensor), $R$ its trace and $G$ is  the gravitational Newton's constant. The energy-momentum symmetric tensor $T_{\mu\nu}$ depends on the fields, their covariant derivatives and the metric. Greek labels refer to four dimensional spacetime and we use metric signature (-+++).



%



%

\section{Tachyons}

The subject of tachyons, even if still speculative, may deserve some attention for reasons that can be mention as the larger scheme that one tries to build up in order to incorporate spacelike objects in the Special Relativity theory. The superluminal classical objects can have a role in elementary particle interactions and right even in astrophysics and it might be tempting to verify how far one can go in search of behavior and the nature at a classical level of the dark energy just by taking account of the possible existence of faster-than-light classical particles. 
Although so far there is no definitive evidence of the existence of faster-than-light particles, the Special Relativity allows the possibility of spacelike 4-velocities. Some researchers have ventured to give a physical theoretical basis to the tachyons \cite{Ehrlich,Bilaniuk,Feinberg,Recami_1,Recami_2,Recami_3}. 

The energy-momentum $p^{\mu}$ in Special Relativity theory can drive three possible equations,
\begin{equation}
 \label{bradyons}
 p_{\mu}p^{\mu} = - m^{2} < 0 \hspace*{1cm}\mbox{timelike momentum,}
\end{equation}
\begin{equation}
 \label{luxons}
 p_{\mu}p^{\mu} =  0 \hspace*{2.8cm}\mbox{lightlike momentum,}
\end{equation}
\begin{equation}
 \label{tachyons}
 p_{\mu}p^{\mu} = + m^{2} > 0 \hspace*{1cm}\mbox{spacelike momentum.}
\end{equation}
The timelike momentum, equation (\ref{bradyons}), is used to describe the energy-momentum of ordinary massive particles with subluminal velocities $v<1$. The subluminal particles are named by ``bradyons'' (as opposed to tachyon) \cite{Cawley}.
The lightlike momentum, equation (\ref{luxons}), is used to describe the energy-momentum of massless particles that travel exactly at the speed of light $c = 1$, such photons and are named by ``luxons''. As is well known, the superluminal particles, with superluminal velocities $v>1$, have been given the name ``tachyons'' coined by G. Feinberg \cite{Feinberg}.  Their  energy-momentum equation is the spacelike momentum equation (\ref{tachyons}).

We will start from the fact that relativistic energy and three-momentum are given by
\begin{equation}
 E = \gamma m  \hspace*{1cm} \mbox{and} \hspace*{1cm} |{\bm p}| = \gamma m |\bm{v}|.
\end{equation}
Thus we have that for timelike momentum, the equation (\ref{bradyons}) yelds the known Lorentz factor given by 
\begin{equation}
 \gamma = \frac{1}{\sqrt{1-v^2}} \hspace*{1cm} \mbox{for bradyons.} 
\end{equation}
Though, for spacelike momentum, the equation (\ref{tachyons}) yelds another format for Lorentz factor 
which is given by

\begin{equation}
\label{Lorentz_Factor_Tachyons}
 \gamma_{_{T}} = \frac{1}{\sqrt{{v^2}-1}}, 
\end{equation}
valid only for tachyons with superluminal velocities $v>1$ \cite{Feinberg}.
As consequences of this spacelike momentum, the limits
\begin{equation}
 \lim_{v \rightarrow 1}  \gamma_{_{T}} = \infty \hspace*{1cm} \mbox{and} \hspace*{1cm}  \lim_{v \rightarrow \infty}  \gamma_{_{T}} = 0,
\end{equation}
imply that the energies of tachyons, $E = \gamma_{_{T}}m $,  are infinite when their velocities $v \rightarrow 1$  and on the other hand, the energies of tachyons are zero  when their velocities are infinite. Regarding the three-momentum, the tachyons have modulus infinite of momenta when their velocities  $v \rightarrow 1$ and have modulus of momenta $|{\bm p}| = m$ when their velocities are infinite. Such discussions are well formalized in the references \cite{Recami_1,Recami_2,Recami_3}.
From equation (\ref{tachyons}) the spacelike 4-momentum for tachyons has been defined as $p^{\mu} = m U^{\mu}$ where $U^{\mu}$ is the spacelike 4-velocity. Then it follows that,
\begin{equation}
\label{quadrado_velocidade}
 U_{\mu} U^{\mu} = + 1 \hspace*{1cm} \mbox{for tachyons.}
\end{equation}
Many authors have investigated the possibility of describing particles that travel faster than light within the special theory of relativity. A review about this can be found in the reference \cite{Ehrlich}.
G. Feinberg in reference \cite{Feinberg} had argued  that the objections generally raised to the existence of such particles are not valid in a relativistic quantum theory. In the references \cite{Recami_1, Recami_2, Recami_3} there are discussions about the phenomenology, kinematics, causality and interactions concerning of tachyons.
Arguing in the Feinberg's own words \cite{Feinberg}, ``the limiting velocity is $c$ but a limit has two sides''.

\section{The tachyon dust energy-momentum tensor}

For researchers who had worked with the topic of superluminal particle phenomenology, in an effort to explain the interaction of these particles in the context of General Relativity, they reasonably stated that the energy-momentum tensor of a tachyon dust \cite{Foster, Schwartz_1, Singh,Srivastava} is given by
\begin{equation}
\label{dust_tensor_1}
 T_{\mu\nu} = \rho \, U_{\mu} U_{\nu},
\end{equation}
where $\rho$ is the energy density of the assemblage of dust tachyon and $U_{\mu}$ is the spacelike 4-velocity of the dust.
In fact for ordinary matter, the bradyons, the dust energy-momentum tensor has the same algebraic format of equation (\ref{dust_tensor_1}) with timelike 4-velocities. It is important to point out that luxons, with lightlike 4-velocities, have the dust energy-momentum tensor in a similar algebraic format given by $ T_{\mu\nu} = \rho \, k_{\mu} k_{\nu}$ where $k_{\mu}k^{\mu} =0$. This energy-momentum tensor is called dust null of pure radiation field  \cite{Kramer}. 
Let us emphasize that the photons (luxons) move on null lines with null time proper and consequently the 4-velocity can not be defined, or in other words, there is no frame in which light is at rest, so there is no comoving reference frame for photons (luxons). It follows that for an observer in the matter world of bradyons it is hard to define a comoving reference frame for tachyons. However in contrast  we can analyse the congruence of spacelike curves in spacetime in accordance with authors in the references \cite{Bilaniuk,Feinberg,Recami_1,Recami_2,Recami_3}. 
In the case of subluminal particles it is possible to define the dust energy-momentum tensor in a  comoving reference frame where we have the rest-mass density $\rho = n_{0} m$, such that $n_{0}$ is the number of particles of mass $m$ per unit volume in a comoving reference frame. If we consider a box with volume, $V_{0}=\Delta x_{0}\,\Delta y_{0}\, \Delta z_{0}$, containing the particles of mass $m$ in a comoving reference frame, the density number in other frame in which the particles are not at rest must be $n = n_0\gamma$ due to length contraction in the direction of motion. Assuming this principle valid for superluminal particles, the density number for tachyons has the similar algebraic form,
\begin{equation}
\label{density_number}
 n_{T} = \frac{n_{0}}{\sqrt{v^2 -1}}
\end{equation}
where $n_{0}$ can be the measured density number in a comoving reference frame of superluminal particles with speed $v>1$. Must be observed carefully when the speed tend to the speed of the light, the density number $n_{T} \rightarrow\infty$ similar to the subluminal particles. But when the speed of supeluminal particles tend to the infinite speed, the density number $n_{T} \rightarrow 0$. This phenomenon can be interpreted as an expasion of the volume that contain the tachyons. For bradyons the length is contracted when the speed varies from zero to the speed of the light, while for tachyons have the length contracted when  the speed varies from infinite speed to the speed of the light, or in other words, for tachyons the length observed by us does expand with increasing speed, thus the measured length for a ruler made of tachyons will have infinite length when the velocity will be infinite. Tachyons with infinite speed must travel an infinite distance in a time interval contracting to zero. Then we can define the dust energy-momentum tensor for tachyons in the similar way the definition of the  dust energy-momentum tensor of ordinary matter as the tensorial product $\bm{T} = \bm{p} \otimes (n_{0}\bm{U})$, where $\bm{p}$ is the 4-momentum and $\bm{U}$ is the 4-velocity, both spacelike vectors and $n_{0}$ is the density number for tachyons. This tensor has components given by the equation (\ref{dust_tensor_1}),  where $\rho=n_{0}m$.
The energy density is 
\begin{equation}
 T^{00} = \frac{n_{0}m}{v^2 -1} = n_{T}E,
\end{equation}
where $E = m\gamma_{_T}$.
When $v$ tends to light speed the energy density $T^{00} \rightarrow\infty$ and when the $v$ tends to infinite speed the  energy density $T^{00} \rightarrow 0$ as expected for tachyon energies. The $i$ momentum density and the energy flux across $x^{i}$ surface are given by
\begin{equation}
 T^{i0} = T^{0i} = \frac{n_{0}m v^{i}}{v^2 -1} = n_{T}p^{i}.
\end{equation}
When $v$ tends to light speed the momentum density and the energy flux are infinite and they are null when the  $v$ tends to infinite speed. Finally, the  flux of $i$ momentum across $x^{j}$ is given by
\begin{equation}
 T^{ij} = \frac{n_{0}m v^{i}v^{j}}{v^2 -1} = n_{T}p^{i}v^{j}.
\end{equation}
Without question, these flux are infinite when $v$ tends to the light speed. The $T^{ij}$ tend to finite values when the velocity of the dust tends to the infinite speed. In term of the direction cosines $v^{i} = v\cos(\alpha^{i})$, then in the limit $v\rightarrow\infty$ we have that  $T^{ij} = \rho \cos(\alpha^{i}) \cos(\alpha^{j})$. 
One can reassert that the genuine equation of tachyon dust energy-momentum tensor is the equation (\ref{dust_tensor_1}) and this equation is the one must connect with Einstein's equation (\ref{Einstein Equation}) when one wants to study the interactions between dust tachyon and gravitational field \cite{Foster, Schwartz_1, Singh, Srivastava}.


\section{Tachyon perfect fluid in General Relativity} 

In General Relativity, a curved spacetime with metric tensor $g_{\mu\nu}$, the  tangents to the timelike congruence yelds a vector field. The timelike tangents are often taken to be the 4-velocity of an  ordinary fluid. For the tachyon fluid we can consider the spacelike congruence with tangents $U^{\mu}$ which must satisfy $U^{\mu}U_{\mu} = +1$.
Similarly to the case of ordinary bradyon fluid, the covariant derivative of a vector field of 4-velocity of the spacelike congruence can be decomposed as follows \cite{Srivastava, Kramer}:
\begin{equation}
 \nabla_{\mu}U_{\nu} = -U_{\mu}\dot{U}_{\nu} + \omega_{\mu\nu} + \sigma_{\mu\nu} + \frac{1}{3}\Theta~h_{\mu\nu},
\end{equation}
where the physical quantities of the tachyon fluid are:\\
the acceleration  
$$\dot{U}_{\nu} = U^{\rho}\nabla_{\rho}U_{\nu}$$   
with $\dot{U}_{\nu}U^{\nu}=0$;\\
the rotation or vorticity $$\omega_{\mu\nu} = \frac{1}{2}\left(\nabla_{\mu}U_{\nu} - \nabla_{\nu}U_{\mu} + U_{\mu}\dot{U}_{\nu}- \dot{U}_{\mu}U_{\nu}\right) $$  
with $\omega_{\mu\nu}U^{\nu}=0$;\\
the shear $$\sigma_{\mu\nu} = \frac{1}{2}\left(\nabla_{\mu}U_{\nu} + \nabla_{\nu}U_{\mu} +  U_{\mu}\dot{U}_{\nu} + \dot{U}_{\mu}U_{\nu} \right)  -\frac{1}{3}\Theta~h_{\mu\nu} $$  
with $\sigma_{\mu\nu}U^{\nu}=0$;\\
the expansion scalar is $\Theta = \nabla_{\nu}U^{\nu}$; \\
the tensor $h_{\mu\nu}$ is the projection operator onto the subspace of the tangent space perpendicular to spacelike tangent velocity vector $U^{\nu}$ \cite{Hawking}. This tensor is defined by
\begin{equation}
\label{projection_tensor}
 h_{\mu\nu} = g_{\mu\nu} - U_{\mu}U_{\nu},
\end{equation}
with $g^{\mu\nu}h_{\nu\rho} = {h^{\mu}}_{\rho}$, $h^{\mu\nu}h_{\nu\rho} = {h^{\mu}}_{\rho}$, $h_{\mu\nu}U^{\nu} = 0$ and ${h_{\mu}}^{\mu}=3$.
In fact $h_{\mu\nu}$ can be  regarded as the induced metric on hypersurface whose  vectors $U^{\mu}$ tangents  to the spacelike congruence are orthogonal to this hypersurface.

In General Relativity the energy-momentum tensor is usually taken to be the derivative of the matter Lagrangian with respect to the spacetime metric tensor $g_{\mu\nu}$. It will be by this way that we will calculate the energy-momentum tensor of the tachyon perfect fluid.
The variational principles constructs an action for gravitation field from which the Einstein field equations of General Relativity (\ref{Einstein Equation}) can be derived. The action for gravitation field interacting with a matter field is given by
\begin{equation}
 S = \int_{\cal D} \left( \frac{R}{16\pi G} + {\cal L}_{m}\right) \sqrt{-g}~d^4x,
\end{equation}
where the first term inside the parentheses is the Einstein-Hilbert action and second term is a suitable Lagrangian density for the matter field.
One obtains the Einstein's field equations (\ref{Einstein Equation})  by requiring that above action be stationary under variation of the fields in the interior of a compact four-dimensional region $\cal D$. Thus the energy-momentum tensor of the fields is given by \cite{Hobson}:
\begin{equation}
 T^{\mu\nu} = \frac{2}{\sqrt{-g}} \frac{\delta{\cal L}_{m}}{\delta g_{\mu\nu}}.
\end{equation}
For ordinary matter, the bradyons, the Lagrangian density of matter field is given by:
\begin{equation}
\label{bradion_fluid}
 {\cal L}_{m} = - n_{0} m \sqrt{-g} \hspace*{1cm} \mbox{for bradyon fluid}, 
\end{equation}
where $n_{0}$  can be the measured density number in a comoving reference frame of subluminal particles of mass $m$ with speed $v < 1$.  For any tachyon fluid with superluminal velocity $v>1$ formed by particles of mass $m$ and density number $n_{0}$ is required that the Lagrangian density of matter field to be:
\begin{equation}
\label{tachyon_fluid_01}
 {\cal L}_{m} = + n_{0} m \sqrt{-g} \hspace*{1cm} \mbox{for tachyon fluid}. 
\end{equation}
As one can be seen that the difference between the squared length of the bradyon momentum (\ref{bradyons}) and tachyon momentum (\ref{tachyons}) is the change of negative signal  to the positive sign respectively, one can change the negative signal of the Lagrangian density of bradyon fluid to the positive sign for tachyon fluid. Thus, we can calculate the energy-momentum tensor for tachyon fluid with Lagrangian of equation (\ref{tachyon_fluid_01}), which results in,
\begin{equation}
 \label{tachyon_fluid_02}
 T^{\mu\nu} = \frac{2}{\sqrt{-g}}\left[\left(\frac{\delta (n_0 m)}{\delta g_{\mu\nu}}\right)\sqrt{-g} 
 +n_0 m\left(\frac{\delta \sqrt{-g}}{\delta g_{\mu\nu}}\right)\right]. \nonumber
\end{equation}

Now let us introduce the definition of Carter and Quintana \cite{Carter_1} of pressure and the shear-stress tensor $q^{\mu\nu}$ by
\begin{equation}
 \label{pressure_tensor_01}
 p^{\mu\nu} = p h^{\mu\nu} - q^{\mu\nu},
\end{equation}
where for an isentropic perfect fluid $q^{\mu\nu}$ is vanished and $p = \frac{1}{3}{p_{\mu}}^{\mu}$  noting that $ p^{\mu\nu}U_{\mu} = 0$.

In accordance with authors in references \cite{Carter_1,Carter_2,Marsden},  when $\rho=n_0 m$ is specified by an equation of state of the form $\rho = \rho(h_{\mu\nu})$ in terms of the six independent components of the projection tensor $h_{\mu\nu}$ the correponding equation of state for the six independent components of the pressure tensor $p^{\mu\nu}$ as function of the projection tensor is given by
\begin{equation}
 \label{pressure_tensor_02}
 p^{\mu\nu} = - 2 \frac{\delta(n_0 m)}{\delta h_{\mu\nu}} - (n_0 m) h^{\mu\nu}.
\end{equation}

Let us return to the first term on the right side of the equation (\ref{tachyon_fluid_02}) where we can enlighten the derivative term as 
\begin{equation}
 \frac{\delta (n_0 m)}{\delta g_{\mu\nu}} = \frac{\delta (n_0 m)}{\delta h_{\kappa\lambda}}\frac{\delta h_{\kappa\lambda} }{\delta g_{\mu\nu}} = \frac{\delta (n_0 m)}{\delta h_{\mu\nu}}, \nonumber
\end{equation}
and then with aid of the equation (\ref{pressure_tensor_02}) to write this term as
\begin{equation}
\label{pressure_tensor_03}
 \frac{\delta (n_0 m)}{\delta g_{\mu\nu}} = - \frac{1}{2} p^{\mu\nu} - \frac{1}{2} (n_0 m) h^{\mu\nu}.
\end{equation}

The second term on the right side of the equation is the most common and simple and results in terms of the projection tensor (\ref{projection_tensor}) in
\begin{equation}
 \frac{\delta \sqrt{-g}}{\delta g_{\mu\nu}} = \frac{1}{2}\sqrt{-g}~g^{\mu\nu} =  \frac{1}{2}\sqrt{-g} \left( h^{\mu\nu} + U^{\mu}U^{\nu} \right). \nonumber
\end{equation}
Then we can put this term and the term of equation (\ref{pressure_tensor_03}) in the equation of energy-momentum tensor (\ref{tachyon_fluid_02}) that results in
\begin{equation}
 \label{tachyon_fluid_03}
 T^{\mu\nu} = (n_0 m) U^{\mu}U^{\nu} - p^{\mu\nu},
\end{equation}
where the pressure tensor $p^{\mu\nu} = p h^{\mu\nu} = p(g^{\mu\nu}-U^{\mu}U^{\nu})$ comes to the equation of the energy-momentum tensor with negative signal. Then it can be written as 
\begin{equation}
 \label{tachyon_fluid_04}
 T^{\mu\nu} = (\rho_{T} + p_{T}) U^{\mu}U^{\nu} - p_{T} g^{\mu\nu}.
\end{equation}
It should be noted that if pressure vanishes the energy-momentum tensor reduces to the dust energy-momentum tensor of equation (\ref{dust_tensor_1}). 
So here we emphasize that the sign in the Lagrangian for tachyon perfect fluid (\ref{tachyon_fluid_01}) must be positive in contrast to the bradyon perfect fluid Lagrangian (\ref{bradion_fluid}).

Note that the energy-momentum tensor of the tachyon perfect fluid obeys the dominant energy condition: $T_{\mu\nu}U^{\mu}U^{\nu}=\rho > 0$, even to spacelike velocities and tends to zero when $v \rightarrow \infty$. 
We emphasize that the energy-momentum tensor of ordinary matter given by $T^{\mu\nu} = (\rho_{B} + p_{B}) U^{\mu}U^{\nu} + p_{B} g^{\mu\nu} $, where the vector velocity $U^{\mu}$ is timelike, the second term $p_{B}g_{\mu\nu}$ has positive signal, while the energy-momentum tensor of the tachyon perfect fluid in equation (\ref{tachyon_fluid_04}) the second term $p_{T}g_{\mu\nu}$ has negative signal. The contraction of the energy-momentum tensor of ordinary matter or bradyon fluid perfect results in $T = -\rho_{B} + 3p_{B}$ while the contraction of the energy-momentum tensor of tachyon perfect fluid gives $T = \rho_{T} - 3p_{T}$. From the energy and momentum conservation equation, $\nabla_{\nu}T^{\mu\nu}=0$ applied to equation (\ref{tachyon_fluid_04}), we find,
\begin{equation}
 U^{\nu}\nabla_{\nu}\rho_{T} + (\rho_{T}+p_{T})\nabla_{\nu}U^{\nu}=0
\end{equation}
and
\begin{equation}
 (\rho_{T}+p_{T})\dot{U}^{\mu} - h^{\mu\nu}\nabla_{\nu}p_{T} = 0.
\end{equation}
These equations are very similar to the energy and momentum conservation equations of perfect fluid of bradyon matter \cite{Hawking}. The only difference is tensor $h_{\mu\nu}$, given by equation (\ref{projection_tensor}), since this tensor is regarded as projection operator onto subspace of the tangent space perpendicular to spacelike tangent velocity vector $U^{\mu}$.

Let us consider now a congruence of spacelike line of tachyons in a $x^1$ coordinate direction with velocity $v>c$ whose vector 4-velocity is $U^{\mu} = (U^0, U^1, 0, 0)$. Let us also consider that the metric of spacetime is 
$$ds^2 = g_{00}dt^2 + g_{11}(dx^1)^2+g_{22}(dx^2)^2+g_{33}(dx^3)^2,$$ 
where the metric tensor is diagonal and we denote $g_{00}=-|g_{00}| $. Therefore, we have from the equation (\ref{quadrado_velocidade}) that,
\begin{equation}
\label{norma de velocidade}
 g_{\mu\nu}U^{\mu}U^{\nu} = - |g_{00}|U^0 U^0 + g_{11}U^1 U^1 = 1. 
\end{equation}
A solution for the two above components of the 4-velocity vector is given by
\begin{equation}
\label{U^0}
 U^0 = \frac{1}{\sqrt{|g_{00}|(g_{11}v^2 -1)}}
\end{equation}
and
\begin{equation}
\label{U^1}
 U^1 = \frac{v}{\sqrt{(g_{11}v^2 -1)}}, 
\end{equation}
where we assume that the velocity of the tachyon beam is superluminal, or $v>1$. We must draw attention to when we have $|g_{00}| = g_{11}=1 $,  the above components of the 4-velocity vector  simplify to 4-velocity in Minkowski spacetime with Lorentz factor given by equation (\ref{Lorentz_Factor_Tachyons}). The covariant components of 4-velocity vector are,
\begin{equation}
\label{U_0}
 U_0 = \frac{-\sqrt{|g_{00}|}}{\sqrt{(g_{11}v^2 -1)}}
\end{equation}
and
\begin{equation}
\label{U_1}
 U_1 = \frac{g_{11}v}{\sqrt{(g_{11}v^2 -1)}}. 
\end{equation}
Therefore, we can calculate the energy-momentum tensor for a beam of tachyons with above 4-velocity vector in a $x^1$ coordinate direction. From equation (\ref{tachyon_fluid_04}) we have the first component,
\begin{equation}
 T_{00} = (\rho_{T}+p_{T})\frac{|g_{00}|}{\sqrt{(g_{11}v^2 -1)}} - p_{T}(-|g_{00}|), \nonumber
\end{equation}
which can be simplified in the equation below,
\begin{equation}
\label{T_00}
 T_{00} = |g_{00}|\left(\frac{\rho_{T} +p_{T}g_{11}v^2}{g_{11}v^2 -1} \right).
\end{equation}
In the same way we have for the following component,
\begin{equation}
\label{T_11}
 T_{11} = g_{11}\left(\frac{\rho_{T}g_{11}v^2 +p_{T}}{g_{11}v^2 -1} \right),
\end{equation}
and the other two remaining $T_{22} = -g_{22}p_{T}$ and $T_{33} = -g_{33}p_{T}$. Note that when we calculate $T={T_{0}}^{0}+{T_{1}}^{1} + {T_{2}}^{2} + {T_{3}}^{3}$ results in $T= \rho_{T} - 3p_{T}$.

The term $g_{11}v^2 -1$ of the equations (\ref{U^0}) and (\ref{U^1}) should be observed more carefully. As we discussed in section II, the tachyons have higher energy when $v\rightarrow 1$ and the energy tends towards zero when $v\rightarrow \infty$. Let us consider high-energy tachyons, with $v \gtrsim 1$ and the situation in which $g_{11}<1$ such that $g_{11}v^2 -1<0$. For this situation we need to review the equations (\ref{U^0}) and (\ref{U^1}) where we obtain,
\begin{equation}
\label{U^0a}
 U^0 = \frac{1}{\sqrt{|g_{00}|(-1)(1-g_{11}v^2)}}=\frac{-i}{\sqrt{|g_{00}|(1-g_{11}v^2)}}
\end{equation}
and
\begin{equation}
\label{U^1a}
 U^1 = \frac{v}{\sqrt{(-1)(1-g_{11}v^2)}} = \frac{-i~v}{\sqrt{(1-g_{11}v^2)}}. 
\end{equation}
This complex reformulation above satisfies and solves the equation (\ref {norma de velocidade}) for situation $1- g_{11}v^2 > 0$.
Then the components of the energy-momentum tensor in the equations (\ref{T_00}) and (\ref{T_11}) change to the form,
\begin{equation}
\label{T_00a}
 T_{00} = -|g_{00}|\left(\frac{\rho_{T} +p_{T}g_{11}v^2}{1-g_{11}v^2 } \right)
\end{equation}
and
\begin{equation}
\label{T_11a}
 T_{11} = -g_{11}\left(\frac{\rho_{T}g_{11}v^2 +p_{T}}{1-g_{11}v^2} \right),
\end{equation}
and the other two remaining $T_{22} = -g_{22}p_{T}$ and $T_{33} = -g_{33}p_{T}$.
A situation where the above results can be applied will be discussed below.


\section{The presence of tachyon perfect fluid energy-momentum tensor in the Friedmann-Lemaître equations}  

Let us consider the spacetime dynamics of the universe described by the Robertson-Walker metric \cite{Hawking},
\begin{equation}
\label{Robertson_Walker_metric}
 ds^2= -dt^2 +a(t)^2\left[\frac{dr^2}{1-kr^2} +r^2(d\theta^2 + \sin\theta~d\phi^2)\right],
\end{equation}
For this metric we  have the following components for the Ricci tensor,
\begin{eqnarray}
 R_{00} &=& -3\frac{\ddot{a}}{a},\cr\cr
 R_{11} &=& \frac{1}{1-kr^2}\left(a\dot{a} + 2\dot{a}^2 +2 k \right),\cr\cr
 R_{22} &=& r^2\left(a\dot{a} + 2\dot{a}^2 +2 k \right),\cr\cr
 R_{33} &=& r^2\sin^2\theta\left(a\dot{a} + 2\dot{a}^2 +2 k \right),
\end{eqnarray}
with curvature scalar,
\begin{equation}
 R= 6\left( \frac{\ddot{a}}{a} + \frac{\dot{a}^2}{a^2} +\frac{k}{a^2}\right).
\end{equation}
With these results above, we can solve Einstein's field equation (\ref{Einstein Equation}), so that we obtain the following equations,
\begin{equation}
\label{Friedmann_1}
 3\left( \frac{\dot{a}^2}{a^2} +\frac{k}{a^2} \right) -\Lambda = 8\pi G ~ T_{00},
\end{equation}
\begin{equation}
\label{Friedmann_2}
 - \frac{2\ddot{a}}{a} -\left( \frac{\dot{a}^2}{a^2} +\frac{k}{a^2} \right) +\Lambda = 8\pi G \left(\frac{1-kr^2}{a^2}\right) T_{11},
\end{equation}
\begin{equation}
\label{Friedmann_3}
 - \frac{2\ddot{a}}{a} -\left( \frac{\dot{a}^2}{a^2} +\frac{k}{a^2} \right) +\Lambda = 8\pi G \left(\frac{1}{a^2r^2}\right) T_{22},
\end{equation}
and 
\begin{equation}
\label{Friedmann_4}
 - \frac{2\ddot{a}}{a} -\left( \frac{\dot{a}^2}{a^2} +\frac{k}{a^2} \right) +\Lambda = 8\pi G \left(\frac{1}{a^2r^2\sin^2\theta}\right) T_{33}.
\end{equation}
We can add the three equations (\ref{Friedmann_2}), (\ref{Friedmann_3}) and (\ref{Friedmann_4}) and substitute the equation (\ref{Friedmann_1}) into that sum to obtain the equation below,
\begin{equation}
\label{Friedmann_5}
 \frac{3\ddot{a}}{a} = \Lambda - 4\pi G \left[T_{00}+ \left(\frac{1-kr^2}{a^2}\right) T_{11} 
  + \left(\frac{1}{a^2r^2}\right) T_{22} + \left(\frac{1}{a^2r^2\sin^2\theta}\right) T_{33}\right].
\end{equation}
Consider that the energy-momentum tensor $T_{\mu\nu}$ is composed of a perfect fluid of ordinary matter, $T_{\mu\nu}^{(B)}$,  and a perfect fluid of energetic tachyons, $T_{\mu\nu}^{(T)}$,  propagating radially at superluminal velocity $v\gtrsim 1$. Also consider a universe of open geometry, $k=-1$ in the metric spacetime (\ref{Robertson_Walker_metric}) and in all above equations.
We can identify $|g_{00}|=1$ and $g_{11} = \dfrac{a(t)^2}{1+r^2}$.
The scale factor $a(t)$ has a relationship with observed redshift $z$ given by,
$ \dfrac{a(t_0)}{a(t_{\mbox{em}})}= 1+z$,
where $a(t_0)=1$ for present time and $t_{\mbox{em}}<t_0$ is the time when the light was emitted by the source \cite{Cheng}. Therefore we have,
${a(t)} = \dfrac{1}{1+z}$.
It can be stated that $a(t)<1$ for $t<t_0$, which it implies that 
 $g_{11} = \dfrac{a(t)^2}{1+r^2}<1$. For a tachyon fluid with high-energy ($v\gtrsim 1$) and for large values of $r$ we have $g_{11}v^2 = \dfrac{a(t)^2v^2}{1+r^2}<1$. We can then make use tachyon fluid whose components are given by the equations (\ref{T_00a}) and (\ref{T_11a}) and $T_{22} = -g_{22}p_{T}$ and $T_{33} = -g_{33}p_{T}$ added together with the bradyon fluid in the equation (\ref{Friedmann_5}) which produces the following result,
\begin{equation}
 \frac{3\ddot{a}}{a} = \Lambda - 4\pi G(\rho_B + 3p_B) + 4\pi G\left[ \left(\frac{\rho_T + p_T\frac{a^2 v^2}{1+r^2}}{1-\frac{a^2v^2}{1+r^2}} \right)
  + \left( \frac{ \rho_T\frac{ a^2 v^2}{1+r^2}+p_T}{1-\frac{a^2v^2}{1+r^2}} \right) + 2p_T\right],\nonumber
\end{equation}
which simplifies into
\begin{equation}
 \label{Friedmann_6}
  \frac{3\ddot{a}}{a} = \Lambda - 4\pi G(\rho_B + 3p_B) + 4\pi G\left[\rho_T\left( \frac{1+ \frac{ a^2 v^2}{1+r^2}}{1-\frac{a^2v^2}{1+r^2}}\right) +p_T\left( \frac{ 3 - \frac{ a^2 v^2}{1+r^2}}{1-\frac{a^2v^2}{1+r^2}}\right)\right].
\end{equation}
The above equation is the Friedmann-Lemaître equation with the presence of tachyon perfect fluid energy-momentum tensor. For long values of $r$ the above equation can be reduced to,
\begin{equation}
\label{Friedmann_7}
 \frac{\ddot{a}}{a} = \frac{\Lambda}{3} - \frac{4\pi G}{3}\left(\rho_B + 3p_B\right) +  \frac{4\pi G}{3}\left(\rho_T + 3p_T\right).
\end{equation}
From the Robertson-Walker spacetime we obtain the acceleration of the scale factor of the universe. The accelerated expansion is obtained when the right-hand side of the equation above is positive. 
The $\Lambda$CDM model states that cosmological constant, $\Lambda$, associated to a vacuum energy or dark energy in empty space, is the main explanation for the contemporary accelerating expansion of space against the attractive effects of gravity. For this model, the cosmological constant contributes with a negative pressure that causes accelerating expansion.
If we consider a cosmological perfect fluid composed by bradyon and tachyon matter the acceleration equation of the scale factor is the result above.
Note that in the presence of energetic tachyon matter ($v\gtrsim 1 $), the Robertson-Walker spacetime (\ref{Robertson_Walker_metric}) in an open geometry, $k=-1$, results in the term $\rho_T + 3p_{T}$ in the equation (\ref{Friedmann_7}) which may contribute to the positive acceleration of the scaling factor.

\section{Conclusion}

In a cosmological framework, the astrophysical issues to explain the accelerating expansion of the universe have pointed to the existence of negative pressure. This phenomenon has found in the astroparticle research many proposals for the particles that make up matter and dark energy of the universe \cite{Gondolo}. Among several possibilities, one should not ignore the possibility of the existence of dust or perfect fluid of tachyons as a component of dark energy and dark matter \cite{Davies}.
As reviewed in this research, a tachyon perfect fluid gives rise a  energy density and a pressure which may contribute to the positive acceleration of the scaling factor of the Robertson-Walker spacetime. If by chance the theoretical assumption of a tachyon fluid is considered, this tachyon fluid may contribute to discussions about negative pressure contributing to the accelerated expansion of the universe.




\end{document}